\documentstyle[psfig]{mn}

\newif\ifAMStwofonts
\AMStwofontstrue

\title{Mode Switching and Subpulse Drifting in PSR~B0826$-$34}
\author[A. Esamdin et al.]
{
A. Esamdin$^{1,2,4,5}$, 
A. G. Lyne$^{2}$,
F. Graham-Smith$^{2}$,
M. Kramer$^{2}$,
\newauthor
R. N. Manchester$^{3}$,
X. Wu$^{1,4}$ 
\\
        $^1$Urumqi Observatory, National Astronomical Observatories, CAS, 40 South Beijing Road,
             Urumqi 830011,China\\
        $^2$Jodrell Bank Observatory, Manchester University, UK\\
        $^3$Australia Telescope National Facility, CSIRO, P.O. Box 76, Epping
             NSW 1710, Australia\\
        $^4$Department of Astronomy, Peking University, Beijing 100871, China\\
        $^5$Chinese Accademy of Sciences - Peking University Beijing Joint
             Astrophysics Center, Beijing 100871, China\\}

\date{Accepted 0000 December 00.
      Received 2004 December 14;
      in original form 2004 October 11}

\pagerange{\pageref{firstpage}--\pageref{lastpage}}
\pubyear{2004}
\begin{document}

\maketitle

\label{firstpage}

\begin{abstract}
We present high-quality observations of PSR~B0826$-$34 at 1374 MHz.
The emission from this pulsar exhibits strong bursts of pulses
followed by long periods of `null' pulses. When it is strong, the
radiation extends through the whole pulse period. We show for the
first time that there is weak emission during the `null' phases, which
should therefore be considered to be a different mode rather than a
null.  During this weak mode the profile is similar to that observed
in the strong mode at low radio frequency.  Using a phase-tracking
method, the pattern of drifting subpulses during the strong mode is
seen to be coherent across the whole profile.  The drift rate is
variable and includes positive and negative values. Thirteen subpulse
bands have been directly observed, covering the whole longitude range.
The subpulses and their spacings ($P_2$) are wider in one half of the
profile than those in the other half.  This difference, and the
variation of observed $P_2$ within the two regions, can be accounted
for if the magnetic pole is inclined to the rotation axis by about
$0.5^\circ$.  These two regions appear to represent radiation from outer
and inner cones.  The intensity modulation of subpulses in all
longitude ranges is related to the magnitude of the drift rate.
\end{abstract}


\section{Introduction}

PSR~B0826$-$34 (PSR~J0828$-$3417) is known as a pulsar whose emission
extends through the whole pulse period ($P = 1.848$~sec). It is a
relatively old pulsar with a characteristic age of $3\times 10^7$yr.
PSR~B0826$-$34 exhibits some extreme forms of subpulse drifting
behaviour, and it switches between strong and weak emission modes. The
weak mode is so weak that it has long been regarded as a complete null
(Biggs et al. 1985, B85 hereafter)\nocite{bmh+85}.

PSR~B0826$-$34 was discovered during the second Molonglo pulsar survey
(Manchester et al. 1978)\nocite{mlt+78}. An early study of this pulsar
by Durdin et al. (1979)\nocite{dll+79} at 408 MHz showed that the
pulsar appears to exhibit pulse nulling for at least 70 per cent of
the time with observed null intervals from a few to 15 000
periods. The drifting subpulses in this pulsar were first observed by
B85.  The integrated pulse profile and most individual pulses extend
over all longitudes of the pulse period, implying that this pulsar is
an almost aligned rotator (i.e. the inclination angle between the
rotation and magnetic axes is very small), and that it is viewed from
a direction close to the rotation axis. At 645 MHz, B85 found 5 to 6
bands of drifting subpulses in part of the profile, and noted that the
drift rate of these bands shows wide variation, including sign
reversals but with no significant variations in the drift-band spacing
($P_2$).

We have recorded two long sequences of individual pulses from PSR
B0826$-$34 at 1374~MHz.  At this frequency, there is strong emission
through most of the pulse period which allows us to investigate the
pattern of drifting subpulses in the strong mode and to trace this
pattern round a complete track in the emitting region.

\section{Observations and data}

The single-pulse observations used in our analysis were made with the
64-m Parkes radio telescope using the central beam of the multibeam
receiver system at 1374 MHz (Manchester et al.  2001)\nocite{mlc+01}.
The receiver system recorded two orthogonal polarizations in 96
channels, each with 3-MHz bandwidth. The channel outputs were delayed
to compensate for the effect of interstellar dispersion, using a
dispersion measure of 65.6 cm$^{-3}$pc, and then summed.

Two long-duration single-pulse sequences were obtained on 2002
Sept 10 and Nov 1. The first sequence (A), lasting 6 hours, was
sampled at 2 millisecond intervals; the second (B), lasting 3.9
hours, was sampled at 0.5 millisecond intervals.

Sequences of several thousand strong pulses, which we note as strong
emission blocks, are typical for this pulsar. Data set A comprised
11670 individual pulses, with about 5900 continuously in a strong
emission block. Data set B comprised 7530 individual pulses, 3900 of
which were continuously in a strong emission block.  These statistics
are similar to those reported by B85.

\section{Integrated pulse profiles}
Fig.~1 shows integrated pulse profiles obtained in the strong mode
at three radio frequencies. The average profile at 1374 MHz was
obtained by integrating all individual pulses within the strong
emission block of data set A. The point of minimum intensity
through the profile was defined as both the zero of longitude and
the zero of intensity. The profile at 606 MHz was recorded in 1998
using the Lovell telescope, and the profile at 408 MHz is taken from
B85. The profiles at two lower frequencies were aligned with the
profile at 1374 MHz by the centroid of the main pulse as described
below.

\begin{figure}
\setlength{\unitlength}{1mm}
\begin{picture}(0,78)
\put(-5,0){\includegraphics{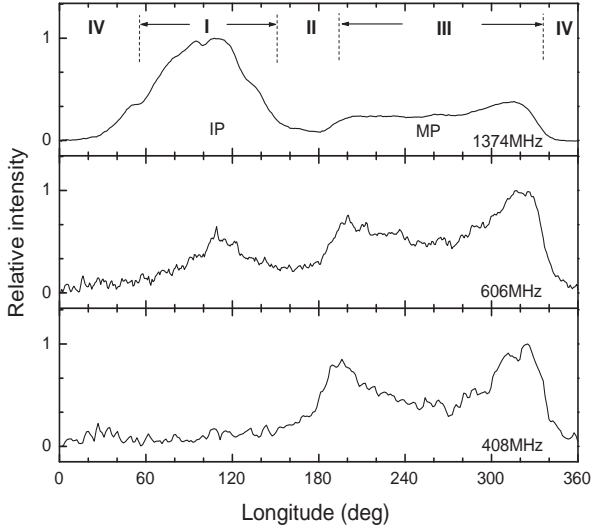}}
\end{picture}
\caption[fig1] {Integrated profiles for the strong mode of PSR
B0826$-$34 at 1374, 606 and 408 MHz, showing the profile evolution
with frequency. The whole longitude is divided into four
regions (I, II, III, IV) according to the difference in emission
and subpulse drifting properties.} \label{fg:threeprof}
\end{figure}

\begin{figure}
\setlength{\unitlength}{1mm}
\begin{picture}(0,78)
\put(-20,85){\includegraphics{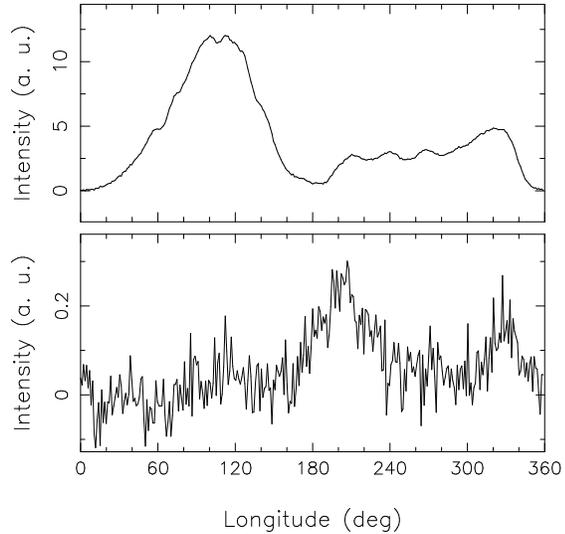}}
\end{picture}
\caption[fig2]{Integrated profiles for the strong (upper panel) and
weak emission mode (lower panel) of PSR~B0826$-$34 at 1374 MHz. Each
profile was obtained from averaging 2048 individual pulses. The units
of intensity are arbitrary, but the same in the two plots and in
Figs.~3, 8 and 9; the lower plot is expanded by a factor of about 45.}
\label{fg:twomode}
\end{figure}

We divide the profile into four regions as shown in Fig.~1.  Regions I
and III, covering longitudes $56^\circ<l<150^\circ$ and
$195^\circ<l<338^\circ$ respectively, display the most obvious
drifting subpulses at 1374 MHz; regions II and IV are two bridge
regions with weaker subpulses.  The components designated as
main-pulse (MP) and inter-pulse (IP) by B85 are within region III and
region I respectively.  Region I is very weak at
408 MHz, but it contains about 24\% and 67\% of the profile energy at
606 and 1374 MHz respectively.

\subsection{The weak mode}

\begin{figure*}
\setlength{\unitlength}{1mm}
\begin{picture}(0,120)
\put(-87,125){\includegraphics{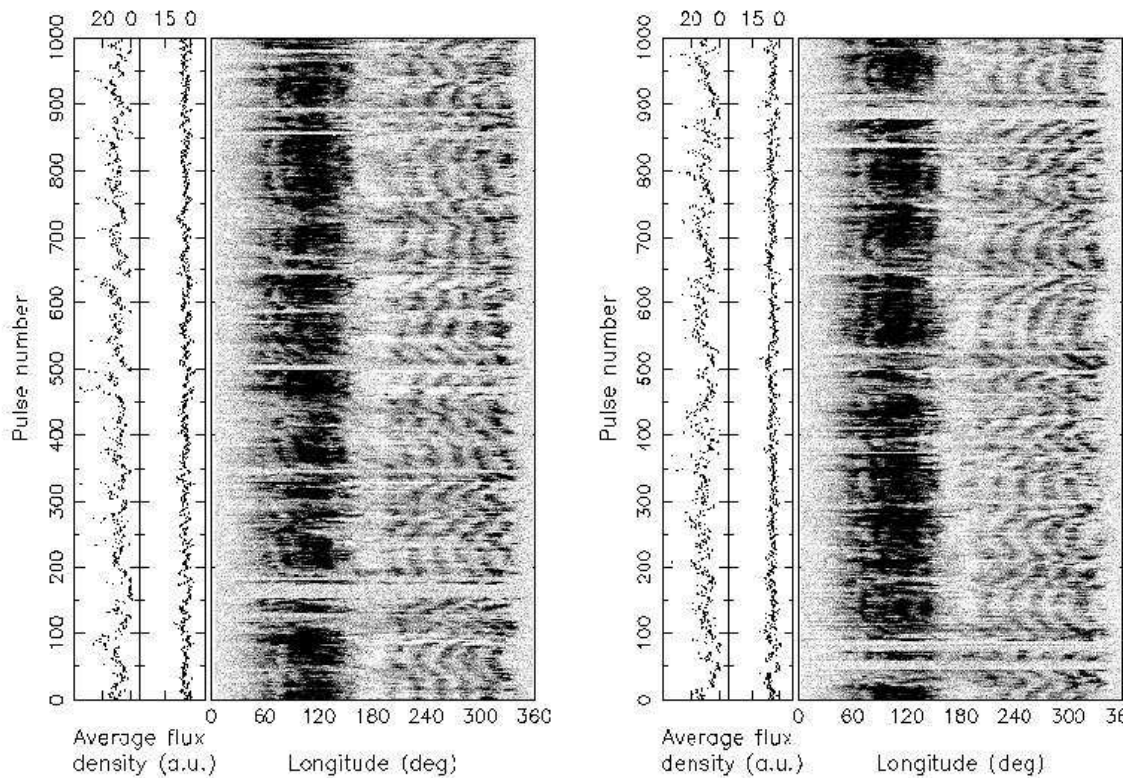}}
\end{picture}
\caption{Two sequences of 1000 individual pulses in strong emission
mode sequences within data sets A (left) and B (right) presented in
grey scale, showing the pattern of subpulse drifting extending through
the whole period. The two panels on the left side of each grey scale
diagram (i.e. longitude-time diagram) show the variation in average
flux densities in regions I (left) and III (right) as a function of
pulse number. The units of flux density are arbitrary, but the same in
all plots and as in Figs.~2, 8 and 9.}
\label{fg:drifting}
\end{figure*}

The weak mode that follows the `normal' or strong mode shown in Fig.~1
generally lasts longer than the strong emission sequences. We have
integrated all pulses received in these apparent `nulls' and find that
the pulsar is in fact detectable at 1374 MHz during this time.  Fig.~2
shows the profiles obtained in the strong and weak modes. The weak
mode profile is found to be similar to this in intensity and shape for
shorter integrations of 1000 periods throughout our observations.

Not only is the mean intensity of the weak mode only about 2\% of that
of the strong mode, but there is also a marked difference between the
two profiles. The pulse profile of the weak mode extends through at
least half of the period, showing a double peak covering region III;
the profile is similar to that of the strong mode at lower frequencies
(Fig.~1). No significant transient effects in pulse phase or intensity
were observed at the transitions between the two modes.

\section{Subpulse drifting}

Fig.~3 presents two sequences of single pulses from both data sets in
the form of a longitude-time diagram. Each contains 1000 successive
pulses in a strong emission sequence. The drifting pattern of
subpulses noted by B85 is seen to extend through the whole pulse
period, although it is indistinct in regions II and IV. 

In this paper, we define the drift rate ($D$) of the drifting
subpulses as $D=\Delta l$ per period ($^\circ/P$), where $\Delta l$ is
the absolute longitude shift in degrees during one pulse period $P$.
A negative value indicates a drift from late to earlier longitudes,
while a positive one indicates a drift from early to later longitudes.

\subsection{Phase tracking and drifting subpulse integration}

Due to the weakness of the subpulses in region II and IV, we
investigate the drift behaviour in these regions most easily in
relation to the subpulse structure in region III, where the subpulse
phase can be determined unambiguously. The positions of subpulses in
regions I, II and IV are then studied relative to those in region III.

The drift phase of the whole pattern was found by extracting the
phases of subpulses within region III by convolution with an
appropriate template in a similar manner to that used by Lyne \&
Ashworth (1983)\nocite{la83} in studying PSR~B0809+74.  This template
was obtained in the first instance by choosing a part of the profile
within longitude range from $220^\circ - 300^\circ$.  The subpulse
phase in each separate profile of the whole series was then found by
determining the position of the peak of the convolution of the
template with the data in this longitude range.  Across this region
the separation of the subpulses is essentially constant and there is
little amplitude variation from subpulse to subpulse, so that the
subpulse phase is well-determined by this method.  An improved
template was then obtained by adding together a few dozen strong
pulses which all had the same phase as the original template and
extracting the appropriate portion of the resulting profile (Fig.~4).

\begin{figure}
\setlength{\unitlength}{1mm}
\begin{picture}(0,70)
\put(-15,75){\includegraphics{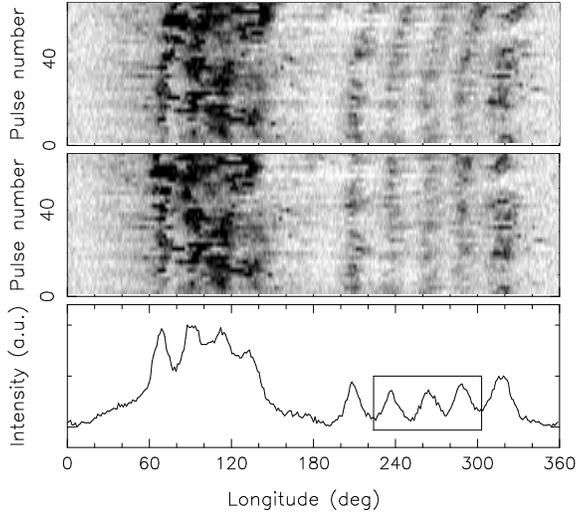}}
\end{picture}
\caption{Construction of a template to follow the drifting phases
of subpulses. The chosen section of profile in longitude range $220^\circ -
300^\circ$ includes three distinct drifting subpulses. The drift
phase of each of a series of 66 successive pulses (shown in the upper
panel) was found by convolution with an initial template obtained from
a single strong pulse. The pulses are again shown in the central
panel, after adjusting their relative phases, and their sum is shown
in the lower panel (the units
of intensity are arbitrary). This template, in the same longitude range, was
used to find the drift phase of all single pulses in the strong
emission mode in both data sets.} \label{fg:template}
\end{figure}

The individual pulses were then divided into ten groups, corresponding
to equal intervals of subpulse phase. All pulses in each of the ten
phase intervals were then co-added, producing the ten profiles shown
in the central panel of Fig.~5.  A coherent pattern of subpulses can
be seen across much of the period, showing that the subpulse
phenomenon is essentially coherent through the whole profile.

\begin{figure}
\setlength{\unitlength}{1mm}
\begin{picture}(0,95)
\put(0,95){\includegraphics{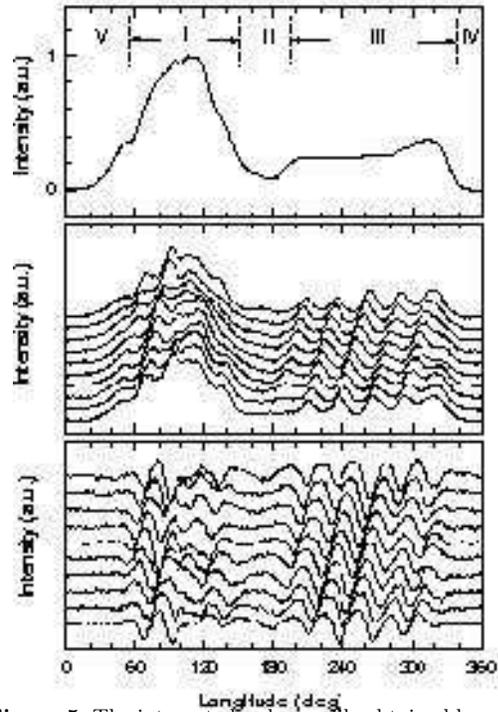}}
\end{picture}
\caption{The integrated pulse profile obtained by adding a total of
9800 individual pulses from the two strong emission sequences (upper
panel), and the integrated profiles obtained by selecting pulses in
ten separate contiguous subpulse phase bins (central panel). The lower
panel of ten profiles was obtained by subtracting the integrated pulse
profile from the ten separate profiles. The units of intensity are
arbitrary and different in the three diagrams.}
\label{tenprof}
\end{figure}

Each of these ten integrated profiles shows the drifting subpulses on
top of a substantial non-drifting contribution to the profile. In
order to improve the visibility of the subpulses, the integrated
profile (top panel in Fig.~5), produced from all single pulses in the
strong mode, was then subtracted from each profile to give the ten
profiles shown in the bottom panel in Fig.~5. These ten profiles show
the integrated subpulse pattern in ten phases of the drift. The ten
profiles are presented in grey-scale format in Fig.~6, repeating twice
in longitude and 14 times in phase, allowing the weak subpulses in
regions II and IV to be followed by eye.  The successful integration
of all subpulses to produce Fig.~6 shows that the pattern is stable
and repeatable over the whole data set.

\begin{figure}
\setlength{\unitlength}{1mm}
\begin{picture}(0,68)
\put(-10,72){\includegraphics{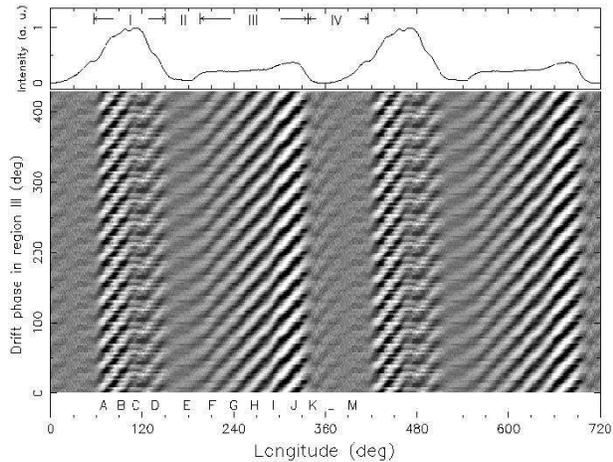}}
\end{picture}
\caption {Integrated drifting subpulse diagram in grey scale (lower
panel). In this presentation the pattern of the ten remnant profiles
in the lower panel of Fig,~5 is repeated twice in longitude and 14
times in subpulse phase, allowing the weak subpulses in regions II and
IV to be followed by eye. 13 drifting subpulses across one pulse
period are labelled A to M. The differences in inclination of the dark
stripes indicates the variation of subpulse spacing in different
longitude regions.  The integrated pulse profile (repeated twice in
longitude) is shown in the upper panel with arbitrary intensity
scale.}
\label {intsubpulse}
\end{figure}

\subsection{The pattern of subpulse drifting}

\begin{figure*}
\setlength{\unitlength}{1mm}
\begin{picture}(0,115)
\put(-88,120){\includegraphics{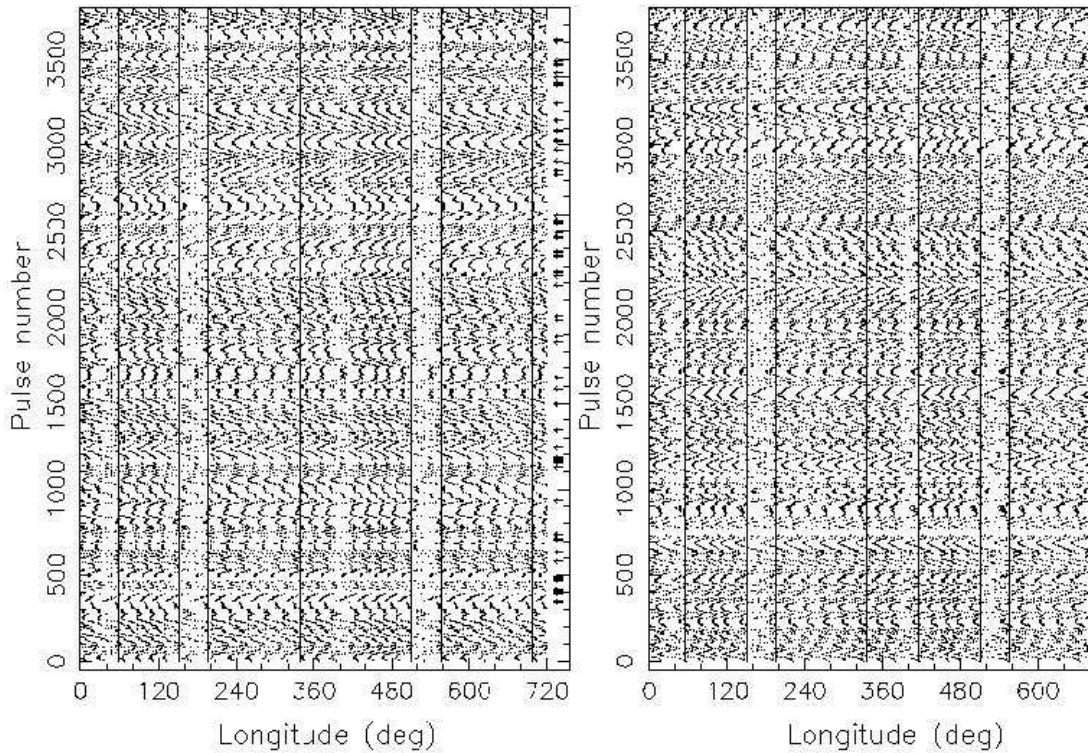}}
\end{picture}
\caption{Drifting patterns from data set A (left panel) and data set
B, each produced by 3800 single pulses, showing the large time scale
drifting property of the pulsar.The positions of drifting subpulses
are presented as dots. The pattern extends twice in longitude to show
more clearly the drifting in region IV. The vertical lines in the
diagrams indicate the four defined longitude regions. The drifting
pattern was interrupted by pulses in weak emission mode,lasting for 1
to 40 periods, indicated by arrows on the right side in both panels.}
\label{dpattern}
\end{figure*}

The drifting subpulses which can be seen through the whole range of
longitude comprise 13 successive drifting sub-bands, labelled A to M
in Fig.~6.  The drifting subpulse pattern continues through the
longitude region around the minimum of intensity in the integrated
profile, confirming that there are regularly drifting subpulses in the
whole longitude range. The least distinct pattern throughout the
diagram appears in the longitude range $20^\circ <l<56^\circ$. (Note,
however, that there is significant integrated-profile energy in this
range and at all longitudes, as shown in the top panel in Fig.~6; this
may not be directly related to the drifting subpulses.)

\subsection{The spacing ($P_2$) and width (FWHM) of drifting subpulses}

As can be seen in Fig.~3, and more clearly in Fig.~6, the longitude
intervals between subpulses ($P_2$) are different in the two prominent
regions I and III, and they vary slightly with longitude. $P_2$ ranges
from $19^\circ$ to $23.5^\circ$ in region I and from $26.8^\circ$ to
$28^\circ$ in region III; in our later discussion we use average
values of $22.2^\circ$ and $27.5^\circ$ for these regions. (B85
reported that the average $P_2$ in region III at 645 MHz is
$29^\circ$).

The full width at half maximum (FWHM) of drifting subpulses in regions
I and III was measured using 20 short sequences of single pulses from
both data sets. Each sequence contained 20 successive single pulses in
a nearly stationary state (i.e. drift rate $D<0.5^\circ$ in either
drift direction). For each sequence, the single pulses were co-added
after adjusting phases of drift to produce one profile. The
subpulses in region I appear to overlap; we therefore fitted
Gaussian curves to obtain the FWHM for each component.

The FWHM of drifting subpulses in region III ranges from $10.7^\circ$
to $14.7^\circ$, with an average of $13.1^\circ \pm 1.9^\circ$. Strong
subpulses tend to be wider, and to occur in the later longitudes of
the region.

The FWHM of subpulses in region I ranges from $16^\circ$ to
$24^\circ$, with an average of $19^\circ \pm 3.5^\circ$.  Strong
subpulses again tend to be wider; they tend to occur around the
middle of the region.

\subsection{The variable drift rate of the whole pattern}

Using the drift phase of each pulse, determined as in Section 4.1, we
plotted the position of each subpulse in all 7600 single pulses of
both data sets as shown in Fig.~7. The short sequences of weak-mode
pulses, within which the drift phase could not be traced, are
indicated by arrows.

The subpulse pattern can drift from early to later longitudes
(positive drift) or from late to earlier longitudes (negative
drift). The transition from negative to positive drift is usually
smooth, while the transition from positive to negative drift is often
abrupt. The most rapid drifting, both positive and negative, occurs
immediately before and after these abrupt transitions, which we refer
to as turning points. At least four abrupt transitions of drift
directions from positive to negative were also noticed by B85,
designated by them as a `runaway' effect, in a sequence of about 2860
single pulses at 645 MHz.  Our data sets suggest that the progress
from turning point to turning point is a repeating pattern, with a
time scale ranging from tens to some hundreds of pulsar periods.

The drift rate in region III ranges from about $-3.2^\circ /P$
(late to early) to about $3.6^\circ /P$ (early to late), changing
continuously within some tens to several hundreds of pulse
periods. This range is significantly larger than that from $-1.5^\circ
/P$ to $2.1^\circ /P$ measured by B85 at 645 MHz; their smaller value
for the range may however be due to the lower signal-to-noise ratio of their
data.

As shown in Fig.~3 and Fig.~7, the average drift rate of drifting
subpulses in region I is slightly smaller than that in region III.  In
region II, as well as in longitude range $20^\circ< l <56^\circ$, the
drifting subpulses present a larger drift rate than others within the
whole longitude range.

\section{Intensity modulation}

B85 noted that strong subpulses in region I occurred more frequently
when the drift was from early to later longitudes. We looked for and
found similar effects in our observations. In Fig.~8, we compare the
integrated profiles produced by single pulses within three drift rate
ranges so as to show the variation of average profile intensity in
three different drift states. The drift ranges we chose to produce
these average profile are $D\geq 0.5^\circ /P$, $D\leq -0.5^\circ /P$
and $|D|<0.5^\circ /P$, measured in region III. Relative to the
average profiles produced by pulses with the most negative drift rates
(solid lines), the profiles corresponding to the small (dashed) and
most positive (dotted) drift rates have increasingly enhanced
intensity in the whole of region I, while the profile around the
leading and trailing edges of region III is weakened.  These
differences are important in our discussion of the possible role of
aliasing in our interpretation of the drifting subpulse pattern.
\begin{figure}
\setlength{\unitlength}{1mm}
\begin{picture}(0,67)
\put(-7,73){\includegraphics{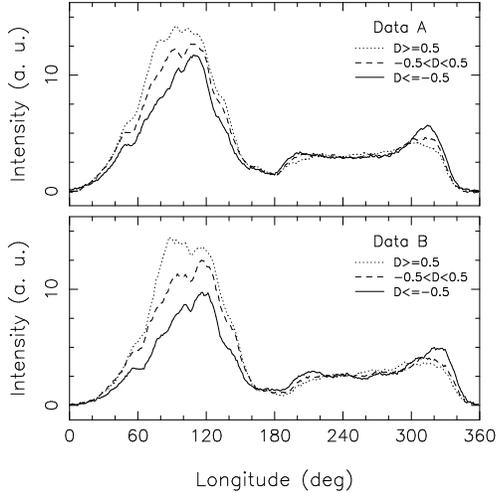}}
\end{picture}
\caption{Integrated profiles produced by single pulses in three drift
rate $(D)$ ranges measured in region III, D $\ge 0.5^\circ$/P (solid
line), $|D|\le 0.5^\circ /P$ (dashed line) and D $\le -0.5^\circ /P$
(dotted line).  The upper and lower panels show the profiles obtained
from data sets A and B respectively.  The units of intensity are
arbitrary, but the same in both panels and as in Figs.~2, 3 and 9.} 
\label{threeprof}
\end{figure}

\begin{figure}
\setlength{\unitlength}{1mm}
\begin{picture}(0,60)
\put(-6,65){\includegraphics{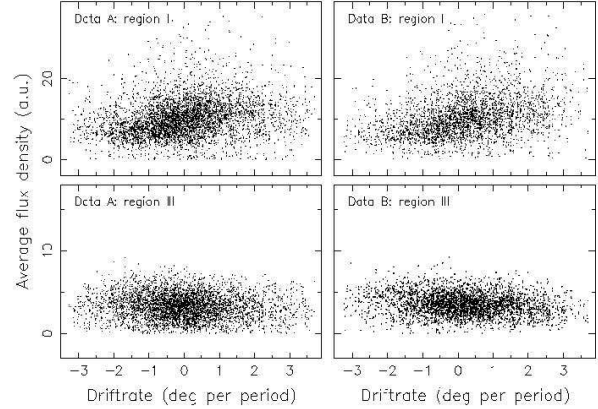}}
\end{picture}
\caption{Average flux densities in region I and III of single pulses
as a function of drift rate. The upper-right and lower-right panels
show the average flux densities in region I and III from data A
respectively. The upper-left and lower-left panels show the average
flux densities in region I and III from data B respectively. The units
of flux density are arbitrary, but the same in all panels and as in
Figs.~2, 3 and 8.}
\label{fluxdens}
\end{figure}
 The dependence of pulse profile on drift rate can also be followed
for individual pulses by integrating the intensity separately over
regions I and III, as shown in Fig.~9. Here the two plots (repeated
for the two dats sets A and B) show the intensities of single pulses
averaged in regions I and III as a function of drift rate.  The
average intensities used in these plots are shown on the left and
right hand sides respectively of the longitude-time diagrams in
Fig.~3.  The modulation of average flux density in these regions is
clearly associated with the drift rate. The two top panels in Fig.~9
(from the two data sets) confirm the conclusion from Fig.~8 that the
emission in region I increases with the increasing drift rate, while
the two lower panels show a small opposite effect in region III.

As we noted in the previous section, the drift behaviour of the pulsar
frequently follows a repeating pattern from turning point to turning
point. Generally, the whole pattern of subpulses drifts from late to
earlier longitudes, is stable for several pulses, and then drifts
from earlier to later longitudes.  Fig.~9 implies that the intensity
modulation of the pulsar follows the same cycle as the
drift. Frequently, subpulse emission in region III rises
significantly in pulses immediately before the turning points, and it
drops significantly at the beginning of the next drifting cycle.

Previous studies in a number of other drifting pulsars have noticed
that the changing of integrated profile shape is associated with a
variation of drift rate. This has been observed in B0031-07 (Huguenin
et al. 1970)\nocite{htt70}, B0809+74 (Lyne \& Ashworth 1983;
Davies et al. 1984; Van Leeuwen et al.
2002)\nocite{la83,dls+84,vkr+02}, B1112+50 (Wright, Sieber \&
Wolszczan 1986)\nocite{wsw86}, B1918+19 (Hankin \& Wolszczan
1987)\nocite{hw87}, B1944+17 (Deich et al. 1986)\nocite{dchr86},
B2319+60 (Wright \& Fowler 1981)\nocite{wf81}, although no reversal of
drift direction is involved in these pulsars.

The variability of observed drift rate of the pulsar, including
reversals of the drift direction, appears to be inconsistent with the
{\bf E}$\times${\bf B} drift of sparks described by Ruderman \&
Sutherland model (Ruderman \& Sutherland 1975)\nocite{rs75}.  Gupta et
al. (2004)\nocite{ggks04} note, however, the possibility that the
observed drift rate is aliased by the rotation rate of the pulsar, in
which case the drift could be in a constant direction with small
variations in rate.  This would require a remarkable coincidence
between the drift rate and the average subpulse spacing, so that the
drift pattern moves by one subpulse interval in one rotation. Our
observations make this interpretation unlikely; the existence of cusps
in the drift pattern, and the difference between the intensity
profiles for positive and negative drift suggest strongly that the
simpler explanation is correct, so that the drift is indeed reversing
its direction of motion.

\section{The geometry of the polar cap}

It has already been pointed out that the wide pulse profile indicates
that this pulsar is observed nearly pole-on, with a small inclination
angle between the magnetic and rotation poles.  The line of sight
follows a circular path round the rotation pole.  We now suggest that
this circular path is within and close to the edge of a polar cap of
emission whose centre is displaced from the rotation pole by a
small angle, as shown in Fig 10. The 13 subpulses
then correspond to a set of 13 emitters spaced uniformly in angle
around the magnetic pole. The whole pattern drifts around the pole in
either direction at a rate of up to 3$^\circ$ to 4$^\circ$ per rotation
period (1.8 sec).

The drift pattern is clearly defined in regions I and III, but in
regions II and IV the drift is more confused.  There is a difference
in subpulse spacing $P_2$ between regions I and III, in which the
average values of $P_2$ are $22.2^\circ$ and $27.5^\circ$
respectively.  We interpret this in terms of the offset of the
circular path of the line of sight from the centre of the polar cap,
as shown in Fig. 10. The difference in subpulse spacing is then
accounted for as a difference in radial distance between the line of
sight and the magnetic pole. The emitting regions extend radially, so
that the line of sight cuts the array of emitters in a circle, radius
$\rho$, centred on the rotation pole R.  The magnetic pole M is at a
small angular distance $\alpha$ from R, so that the angle between the
line of sight and M varies between $\rho+\alpha $ and $\rho-\alpha$.
The observed ratio of subpulse spacing is 1.22, giving

\begin{equation} \frac{\rho +\alpha} {\rho -\alpha}= 1.22
\,\,\, {\rm and \,\, hence} \,\,\,
\alpha = 0.1\rho.
\end{equation}

The geometry can be determined in principle from
measurements of the linear polarisation, but the position angle swing
is complex (Lyne \& Manchester 1988)\nocite{lm88}, and possibly
confused by orthogonal polarisation mode switching.  Single pulse
polarisation observations are required to reconstruct the underlying
mode-corrected position angle swing (Gil \& Lyne
1995)\nocite{gl95}.  The typical value for the diameter of the polar
cap is $\sim13^\circ P^{-1/3}$ (Lyne and Manchester 1988), i.e. $10^\circ$
for B0826-34.  If we assume that the line of sight is close to the edge of
the polar cap, so that the radius $\rho\approx 5^\circ$, we find that the
inclination angle $\alpha=0.5^\circ$.

\begin{center}
\begin{figure}
\setlength{\unitlength}{1mm}
\begin{picture}(0,70)
\put(-5,75){\includegraphics{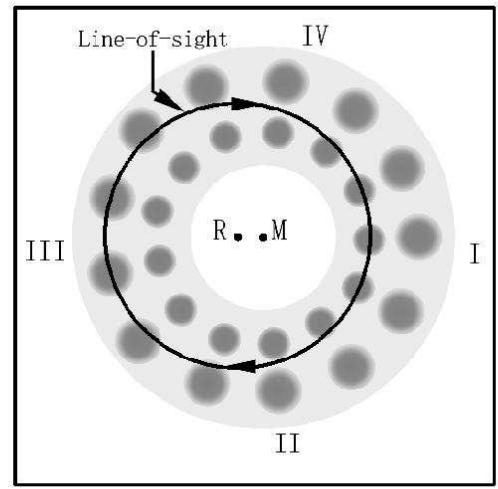}}
\end{picture}
\caption{Map of the polar cap. The subpulse emitters are disposed
uniformly round the magnetic pole M, which is inclined at angle
$\alpha-\beta$  to the rotation axis R.  The line of sight cuts the emitters
in a circle centred on R.  In the two main regions I and III the line
of sight cuts the pattern at two different radial distances from the
magnetic pole; the existence of the two weak regions II and IV suggests
that there may be a radial structure, depicted here as two rings of emitters.}
\label{model}
\end{figure}
\end{center}

The weakness of the subpulses in regions II and IV, and the
poor definition of their tracks in Fig.~6 suggests that there may be a
radial variation of emitter strength. This would accord with the
suggested identification of inner and outer cone components in the
integrated pulse profiles of many pulsars observed at large polar
angles $\alpha$ (Rankin 1983,1993a,1993b; Gil, Kijak \& Seiadakis
1993; Kramer et al. 1994)\nocite{ran83,ran93,ran93b,gks93,kwj+94}.
Such an explanation requires that the conal structure for an almost
perfectly aligned pulsar such as B0826$-$34 is similar to that of much
more oblique rotators.  On the other hand, it is possible that regions
I and III are simply areas of enhanced emissions from bright regions
over a randomly patchy polar region as suggested by Lyne \& Manchester
(1988).\nocite{lm88}

We regard the moding behaviour as a change in a pattern of excitation
underlying the whole of the polar cap; it is not yet possible to say
whether the subpulse drifting continues in the weak mode.

\section{Summary}

The main results of this work are as follows:

1. The `nulls' in the pulses from PSR~B0826$-$34 are not true `nulls'
and the pulsar presents strong and weak emission modes at 1374 MHz. In
its strong emission mode, the radiation extends through the whole
pulse period, and region I is stronger than region III
at this frequency. In the weak emission mode, the profile is similar
to that observed in the strong mode at low radio frequency, at which
region I is the weaker.

2. Using a phase-tracking method, the drifting subpulses of the pulsar
can be traced through the whole pulse period. Thirteen drift
bands have been directly observed, covering the whole longitude range,
and the drifting occurs in both directions.

3. The subpulses in region I and their spacing ($P_2$) are wider
than those in region III.  These differences can be accounted for
if the magnetic pole is inclined to the rotation axis by about
$0.5^\circ$.

4. Regions I and III are consistent with radiation from inner and
outer cones; in regions II and IV of longitude, the line-of-sight
trajectory cuts through a weak emission region between the two cones.

5 .The observed direction of drift is a true reversal within the polar
cap and is not a product of aliasing with the rotation period.

\bibliographystyle{mn}

\bsp

\label{lastpage}

\end{document}